\documentclass[rnote]{aa} 
\def\kms{km\,s$^{-1}$}

\def\hii{H{\sc ii}}
\def\msun{M$_\odot$}

\def\cmdos{cm$^{-2}$}
\def\cmtres{cm$^{-3}$}

\def\radec{RA,Dec.(J2000)}
\def\co{CO(3$-$2)}
\def\tco{$^{13}$CO(3$-$2)}
\def\dco{C$^{18}$O(3$-$2)}
\def\hco{HCO$^{+}$(3$-$2)}
\def\co{$^{12}$CO(3$-$2)}
\def\13co{$^{13}$CO}
\def\18co{C$^{18}$O(3$-$2)}
\def\hcn{HCN(3$-$2)}

\def\hii{H{\sc ii}}
\usepackage{graphicx}
\usepackage{txfonts}
\begin{document} 

   \title{Molecular gas in the star-forming region IRAS\,08589-4714}

   \author{Hugo P. Salda\~no\inst{1}\inst{2},
      J. V\'asquez\inst{2}\inst{3}\inst{4},
      C. E. Cappa\inst{2}\inst{3}\inst{4},
      M. G\'omez\inst{1}\inst{2},  
        N. Duronea\inst{2}\inst{3},
       \and 
        M. Rubio\inst{5}
                  }

   \institute{\inst{1} Observatorio Astron\'omico, Universidad Nacional de C\'ordoba, C\'ordoba, Argentina\\
   \email{hugosaldanio@oac.uncor.edu}\\
   \inst{2} CONICET, Consejo Nacional de Investigaciones Cient\'\i ficas y T\'ecnicas, Argentina\\
   \inst{3} Instituto Argentino de Radioastronom\'ia, CONICET, CCT La PLata, C.C.5, 1894, Villa Elisa, Argentina \\
   \inst{4} Facultad de Ciencias Astron\'omicas y Geof\'isicas, Universidad Nacional de la Plata, Paseo del Bosque s/n, 1900, La Plata, Argentina \\
   \inst{5} Departamento de Astronom\'\i a, Universidad de Chile, Casilla 36, Santiago de Chile, Chile\\
}

   \date{Received September 15, 1996; accepted March 16, 1997}

 
  \abstract
   {}
   {We present an analysis of the region IRAS\,08589$-$4714 with the aim of characterizing the molecular environment.}
   {We observed the 
   $^{12}$CO($3-$2), $^{13}$CO($3-$2), C$^{18}$O($3-$2), HCO$^{+}$($3-$2), and HCN($3-$2) molecular lines in a region 
   of 150$''$ $\times$ 150$''$, centered on the IRAS source, to analyze the distribution and characteristics of the molecular gas
linked to the IRAS source. }
   {The molecular gas distribution reveals a molecular clump that is coincident with IRAS\,08589$-$4714 and with a dust 
   clump detected at 1.2 mm. The molecular clump is 0.45 pc in radius and its mass and H$_2$ volume density are 
   310 $M_{\sun}$ and 1.2$\times 10^{4}$ cm$^{-3}$, respectively. Two overdensities were identified within the clump 
   in \hcn\ and \hco\ lines. A comparison of    the LTE and virial masses suggests that the clump is collapsing in 
   regions that harbor young stellar objects. An analysis of the molecular lines suggests that they are driving molecular outflows. 
} 
   {}

\authorrunning{H. P. Salda\~no et al.}
\titlerunning{Molecular gas toward IRAS\,08589$-$4714}
   \maketitle
%

\section{Introduction}

IRAS\,08589$-$4714 (\radec\ = 09:00:40.5, --47:25:55) {  can be} classified as an ultracompact \hii\ region (UCHII) 
according to the criteria by Wood and Churchwell (1989). {This source coincides with a massive dust clump 
detected in the IR continuum at 1.2 mm by \cite{Beltran_2006}. They estimated a luminosity of 
1.8$\times$10$^{3}$ $L_{\sun}$ and a mass of 40 $M_{\sun}$ for this object.} 

\cite{Wouterloot_1989} detected emission in the $^{12}$CO(1$-$0) molecular line (angular resolution: 43\arcsec) 
toward the IRAS source at $V_{LSR} =$ +5.2 km\,s$^{-1}$. The  molecular line shows an asymmetry in the 
blueshifted peak that is likely produced by noncentral self-absorption and a wing extended toward the red, which is a tracer 
of a potential outflow. \cite{Bronfman_1996} observed the source in the CS(2-1) molecular line at 
$V_{LSR} =$ +4.3 km\,s $^{-1}$, and \cite{Urquhart_2014} detected emission from the high density ammonium 
molecular tracer. The central velocity coincides with that of the CS line. With velocities in the range 4-5 \kms, 
the circular galactic rotation model by \cite{Brand_1993} predicts a kinematical distance of 2.0 kpc. An 
uncertainty of 0.5 kpc is assumed, adopting a velocity dispersion of 2.5 km\,s$^{-1} $ for the interstellar 
molecular gas.

We report molecular line observations of the IRAS source using tracers of low and high density regions 
with the aim of studying the  molecular gas content of the source, identifying dense gas clumps, finding massive YSOs linked to the 
clumps, and identifying possible outflows.

\section{Molecular line observations}

IRAS 08589$-$4714 was observed with the 12 m Atacama Pathfinder EXperiment (APEX) telescope\footnote{Atacama 
Pathfinder EXperiment. APEX  is a collaboration between the Max-Planck-Institut fur Radioastronomie, the European 
Southern Observatory, and the Onsala Space Observatory.}, located in Llano de Chajnantor, in the Puna de Atacama, 
Chile. We carried out observations in the $^{12}$CO(3$-$2), $^{13}$CO(3$-$2), C$^{18}$O(3$-$2), HCO$^{+}$(3$-$2), and  
HCN(3$-$2) molecular lines using the {\it On-The-Fly mapping} technique.  These observations were made on 2014 
June 16, 19, and 21. The CO isotopes were observed with the receiver APEX$-$2 in the spectral range of 
343.8$-$347.8 GHz and in the range 328.6$-$332.6 GHz with a half-power beamwidth  of $\sim$ 18\arcsec, while 
the  HCO$^{+}$ and HCN molecules were observed with the receiver APEX$-$1 in the spectral range of 
265.6$-$269.5 GHz with an HPBW of $\sim$ 22\arcsec\ \citep{Vassilev_2008}. 

\begin{figure}
\includegraphics[width=9cm,height=7cm]{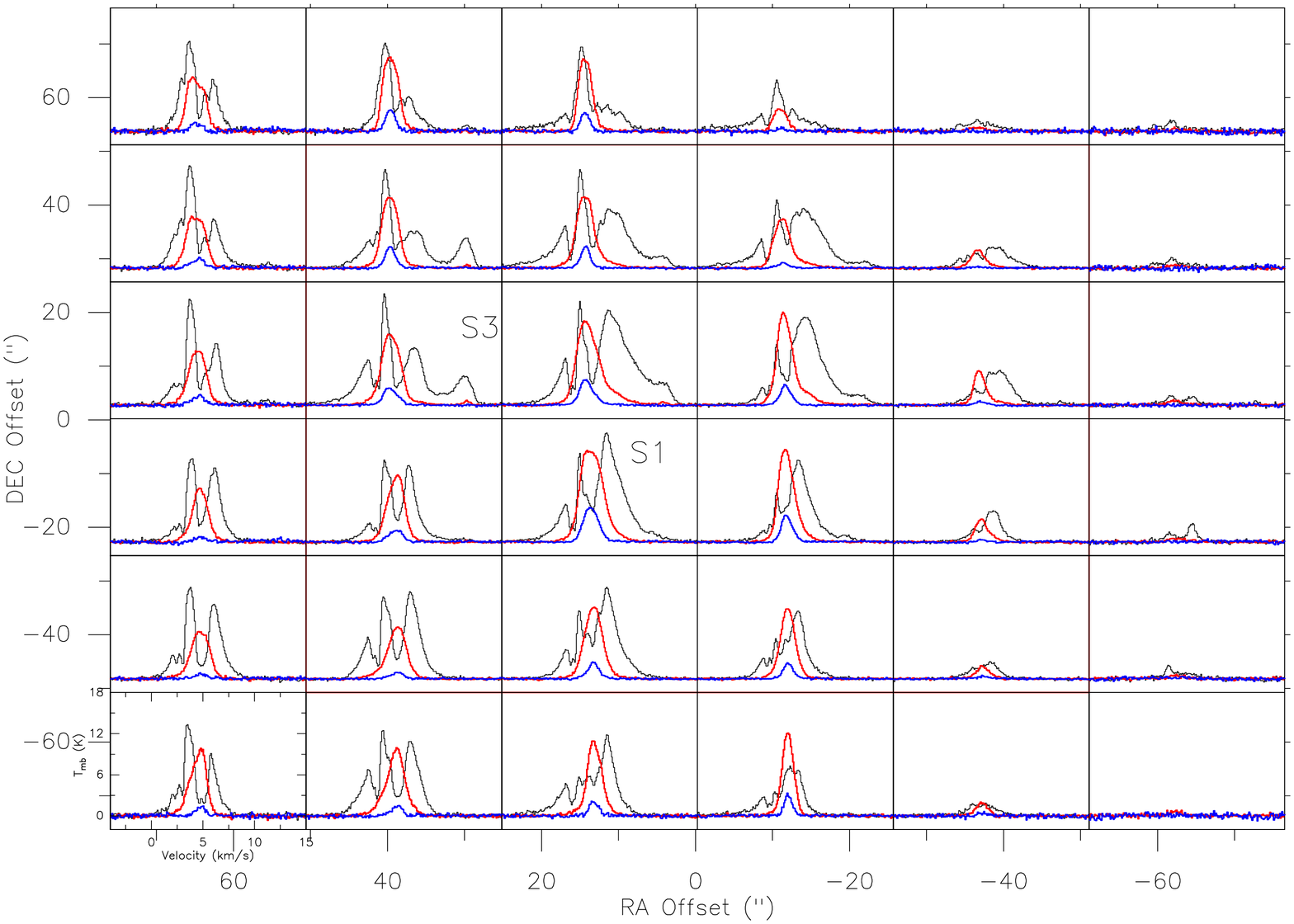}
\caption{{Spectra of the $^{12}$CO(3$-$2), $^{13}$CO(3$-$2), and C$^{18}$O(3$-$2) lines indicated} in black, red, and 
blue, respectively. Each spectrum is the average of the profiles within a field of 
25$''$.5 $\times$ 25$''$.5. The total area is 150$''$ $\times$ 150$''$ for three molecules. 
The locations of sources 1 (S1) and 3 (S3) are labeled. 
In each panel the velocity ranges  from $-$4 to $+$15 km\,s$^{-1}$ and T$_{\rm mb}$ scales from $-$2 to 18 K.}
\label{mapa_perfiles_12_13_18}
\end{figure}

We mapped an area of 150\arcsec $\times$ 150\arcsec\ covering the central region of the IRAS source and the 
1.2 mm emission zone detected by \cite{Beltran_2006}. The APEX$-$1 system temperature is 150 K and 300 K for 
APEX$-$2. The data reduction was performed according to the standard procedure of the CLASS software, 
Gildas\footnote{http://www.iram.fr/IRAMFR/GILDAS/}. The antenna temperature, $T_{\rm A}$, was transformed 
to main-beam brightness-temperature ($ T_{\rm mb} = T_{\rm A} / \eta_{\rm mb} $), using a main beam efficiency 
$\eta_{\rm mb}$ $=$ 0.72 for APEX$-$1 and APEX$-$2 \citep{Vassilev_2008}. 
{  The data were obtained with a velocity resolution of 0.11 km\,s$^ {-1}$. The final {\it rms} is 0.3 K.}

\section{Identification of YSOs}
\label{Identification_of_YSOs}

To investigate the presence of young stellar objects (YSOs) coincident with the IRAS source in the 
{\it Wide-field Infrared Survey Explorer} (WISE) catalog  \citep{Wright_2010}, we applied the following 
criteria \citep{Koenig_2012}. Class I objects are those that satisfy W1 $-$ W2 $>$ 1.0 and W2 $-$ W3 $>$ 2.0,
whereas  Class II objects have W1 $-$ W2 $-$ $\sigma_{1}>$ 0.25 and W2 $-$ W3 $-$ $\sigma_{2}>1.0 $, 
where W1, W2, W3, and W4 are the {magnitudes} in the four WISE bands at 3.4, 4.6, 12, and 22 $\mu$m, 
respectively, and $\sigma_{1}$ and $\sigma_{2}$ are the combined errors of W1 $-$ W2 and W2 $-$ W3, 
respectively\footnote{http://irsa.ipac.caltech.edu/frontpage/}. Three sources with colors of Class I/II 
were identified. Their coordinates and correlation with 2MASS sources are indicated in Table \ref{sources}. 
Sources 1 and 3 are projected onto the dust clump detected by   \cite{Beltran_2006}.

\begin{table}
\centering
\caption{Candidate YSOs in the WISE and 2MASS databases detected toward IRAS\,08589$-$4714}
\begin{tabular}{cccc}
\hline 
\hline
 Source & WISE Source  &  2MASS source & Class \\
\hline
  1 & J090040.97-472601.1 & 09004242-2726050 & I \\
  2 & J090038.59-472648.5 & 09003861-4726489 & I \\
  3 & J090043.08-472539.5 & 09004309-4725394 & II  \\
\hline
\hline
\end{tabular}
\label{sources}
\end{table}

\section{Molecular line analysis}
\label{Molecular_line_analysis}

\subsection{The $^{12}$CO, $^{13}$CO, and C$^{18}$O molecular tracers}\label{spectral_analysis}

Figure \ref{mapa_perfiles_12_13_18} shows the spectra corresponding to the $^{12}$CO(3$-$2), $^{13}$CO(3$-$2), and 
C$^{18}$O(3$-$2) molecular lines (indicated in black, red, and blue colors, respectively).
In this Figure we distinguish two areas in the $^{12}$CO molecular emission: for $\Delta \alpha> -40''$, the emission 
is very complex and shows many components with velocities between $-$3.0 and $+$13 km \,s$^{-1} $, while for 
$\Delta \alpha < -40''$ the emission is weak, below 3 K. 

The profiles of the $^{13}$CO(3$-$2) line reveal two 
distinct regions, as in the case of the $^{12}$CO emission. The emission in the $^{13}$CO(3$-$2) spectra has 
velocities in the range [$+$1, $+$9]\,\,km\,s$^{-1}$ and only one maximum that is centered between $\sim$ $+$4 and 
$+$5 km\,s$^{-1}$. The C$^{18}$O(3$-$2) line profiles are shown in an area smaller (120$''$\,$\times$\,120$''$) 
than in the previous cases. The area of more intense emission has a rather elongated shape from NE to SW. 

\begin{figure*}
\includegraphics[width=9cm,height=8.6cm]{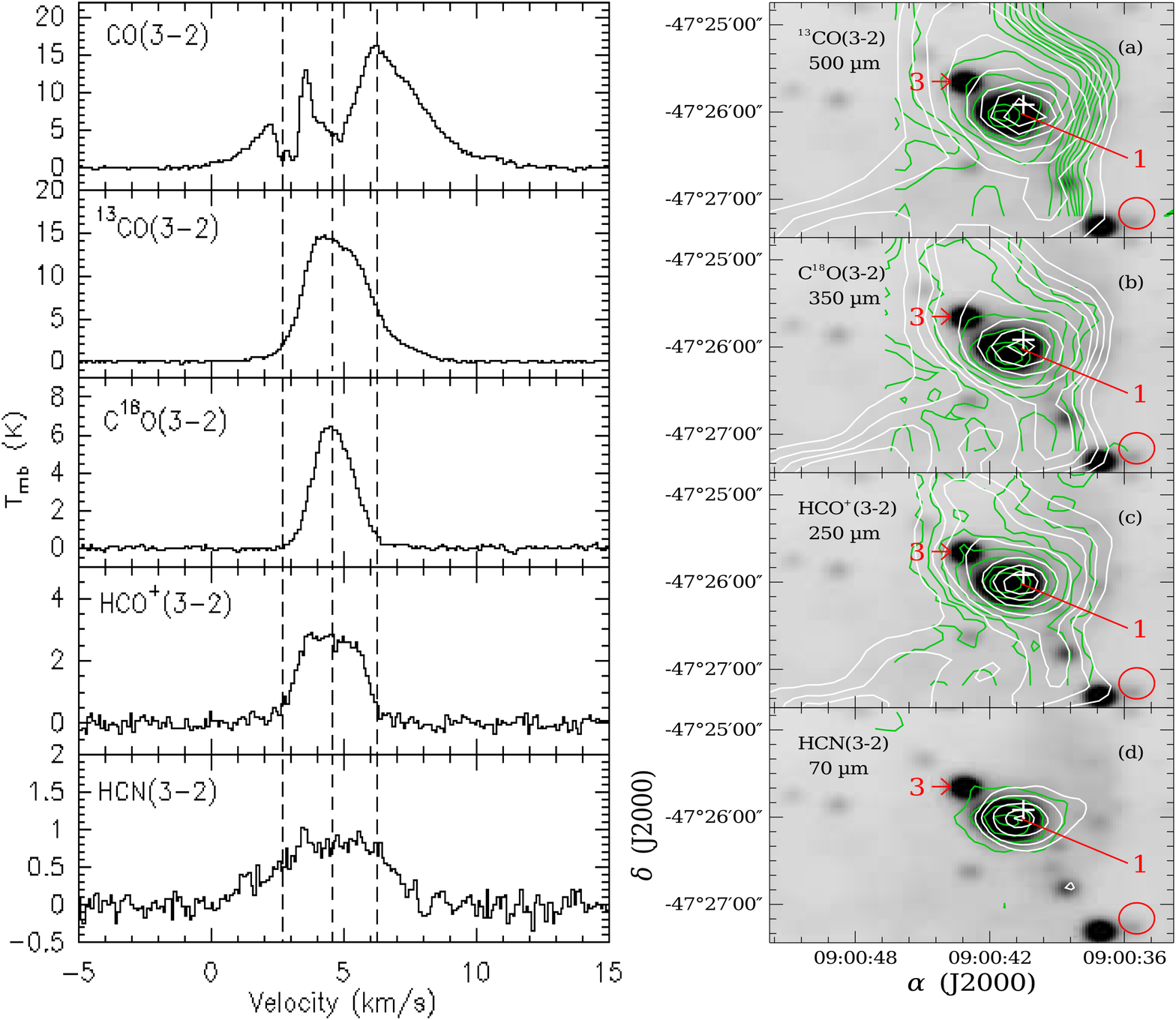}
\includegraphics[width=9cm,height=8.6cm]{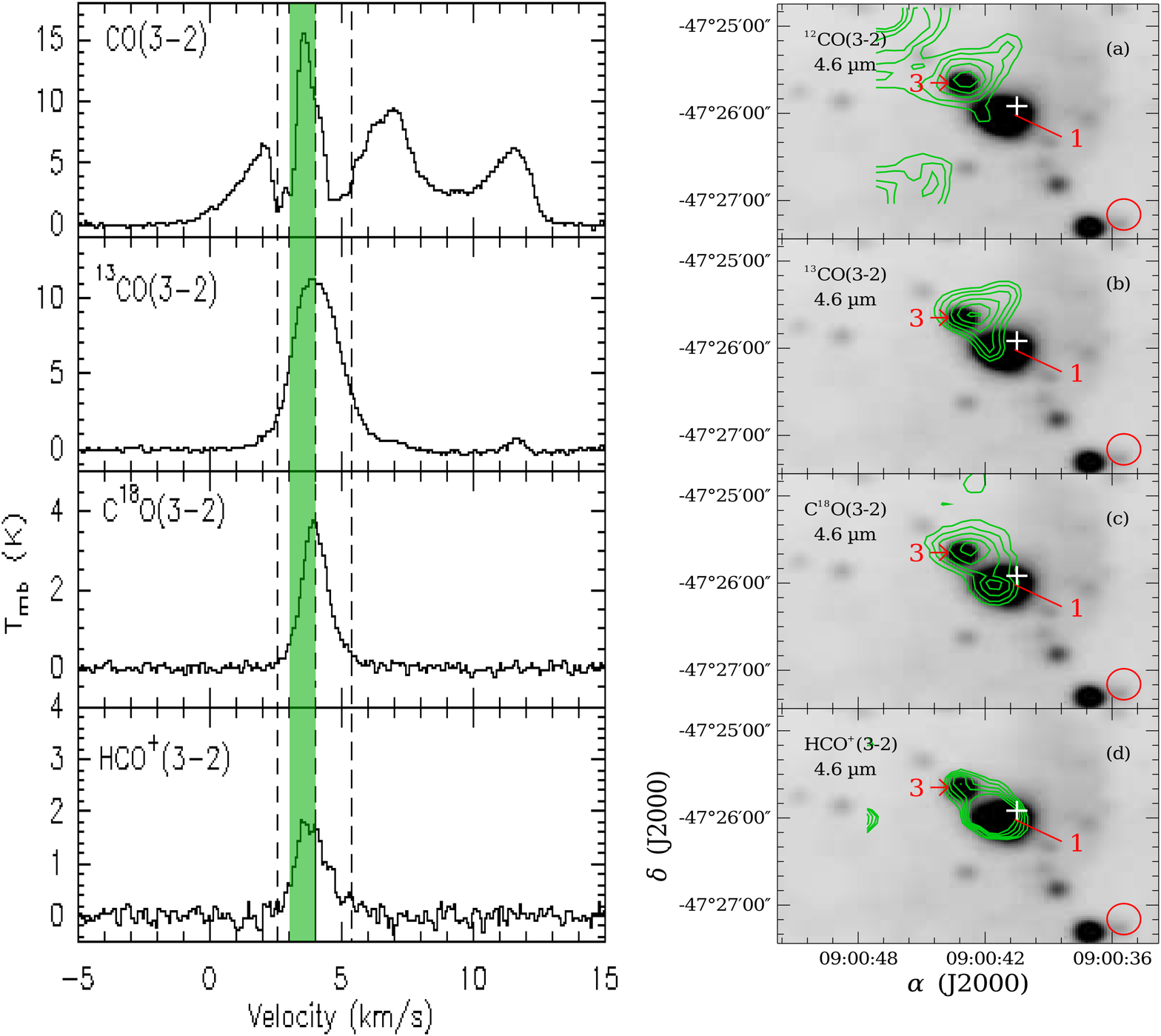}
\caption{{\it Left panels:} Molecular line profiles of $^{12}$CO, $^{13}$CO, C$^{18}$O, HCO$^{+}$, and HCN toward 
source 1. The vertical dashed central line corresponds to $+$4.6 km\,s$^{-1}$. The two remaining vertical dashed 
lines indicate the velocity range [+2.7,+6.3] \kms. Integrated emission maps for the observed molecules (green 
contours) superposed over the 4.6 $\mu$m (grayscale) image. The range of the integrated emission for all molecules 
is [$+$2.7,$+$6.3] km\,s$^{-1}$. The contour levels of the $^{13}$CO are between 4 and 47 K km\,s$^{-1}$.
The contour levels of C$^{18}$O have values between 1.9 and 13.2 K km\,s$^{-1}$
, while the contours of HCO$^{+}$ and HCN are between 0.8$-$7.8 K km\,s$^{-1}$ and 1.1$-$4.1 K km\,s$^{-1} $,
respectively. Herschel maps at 500 (a), 350 (b), 250 (c), and 70 $\mu$m (d) are superimposed in white contours. 
The white cross indicates the IRAS source position, and the red arrows indicate the location of the 
identified sources. The HPBW is shown in the lower right corner of each panel. 
{\it Right panels:}  molecular line profiles toward source 3. The vertical dashed central line corresponds to 
$+$4.0 km\,s$^{-1}$. 
Integrated emission maps of the observed molecules (green contours) superposed onto the 4.6 $\mu$m (grayscale) 
image. The integration range for all molecules is [+3.0,+4.0] km\,s$^{-1}$. The velocity range is indicated 
by the green bar. The contour levels of the CO are between 6.4 and 8.0 K km\,s$^{-1}$.
The contours levels of $^{13}$CO have values between 4.5 and 6.5 K km\,s$^{-1}$,
while the contours 
of C$^{18}$O and HCO$^{+}$ are between 1.2$-$2.3 K km\,s$^{-1}$ and 0.7$-$1.1 K km\,s$^{-1} $, respectively. 
}
\label{molec_cores_1}
\end{figure*}

\subsection{Molecular emission toward sources 1 and 3}

Figure \ref{molec_cores_1}  shows the spectra of the molecular lines observed toward sources 1 and 3 indicated in 
Fig. \ref{mapa_perfiles_12_13_18}. Toward  source 1 {  (see the two panels on the left of Fig. 2)}, the \co\ emission 
consists of three maxima between $\sim$0.0 and $\sim$ $+$12.0 \kms\ that are centered at $\sim$\,\,$+$2.2, $+$3.6, and 
$+$6.3 km\,s$^{-1}$. The profile shows a strong broadening toward more positive velocities and weaker broadening 
toward more negative velocities (all velocities are referred to the LSR). 

The spectra of the remaining molecules display emission between the two external dashed lines 
($+2.7 < {\rm v} <+6.3$\,\kms) with the exception of the \tco\ and \hcn\ lines, which show weak emission 
outside this velocity interval. The \tco\ and \dco\ lines show peak temperatures at the same velocity of a 
depression in the \co\ profile, which is a characteristic shared by the \hco\ and \hcn\ spectra. 

Considering that the emission of \hco\ and \hcn\ corresponds to the densest region of the molecular cloud, 
we adopt ${\rm v}$ $= +$4.6 \kms\ as the systemic velocity, ${\rm v_{sys}}$, for the molecular counterpart 
of source 1 (central dashed line in Fig.\,\ref{molec_cores_1}), in agreement with \cite{Bronfman_1996}.

{The second panel on the left of} Figure \ref{molec_cores_1} {  also shows the spatial distribution of} the 
molecular gas emission in the \tco, \dco, \hco, and \hcn\ lines (green contours), integrated within the 
velocity range enclosed by the two external dashed lines in the {first} panel. The molecular emission 
is superposed onto the 4.6 $\mu$m image (grayscale) and onto Herschel at 500 (a), 350 (b), 250 (c), and 
70 $\mu$m (d) in white contours. The emission of the molecular lines shows a maximum coincident with the 
brightest region in the FIR, and the extension of the emission decreases in size from \tco\ to \hcn. 
The dust emission provides evidence for a similar trend from 500 to 70 $\mu$m. The spatial agreement between the line 
and the continuum emissions reveals the  molecular counterpart of the dust clump. The HCO$+$ and HCN 
emission shows a dense molecular region inside a lower density clump depicted by the $^{12}$CO and 
$^{13}$CO emissions. The molecular emission covers a region with an equivalent radius of 50$''$.

Figure~\ref{molec_cores_1} also shows the molecular line spectra toward source 3 {  (see the two panels on 
the right)}. The \co\ spectrum shows emission between $\sim$ $-$2.0 to $\sim$ $+$13.0 \kms with four peaks. 
The central velocity of the second peak ($\sim$ $+$3.6 \kms) is in agreement with the maximum in the remaining 
molecular spectra ($\sim$ +4.0 \kms), whose emission is enclosed by the two external dashed lines, between 
$+$2.7 and $+$5.3 \kms. 

The  {  spatial distribution of the } emission in the \co, \tco, \dco, and \hco\ lines (green contours) 
toward source 3 is also shown in  Fig.~\ref{molec_cores_1}, integrated in the range [+3.0,+4.0] \kms, 
and overlaid onto the 4.6 $\mu$m image (grayscale). This small velocity interval was chosen to  highlight 
the gas linked to source 3, {  although  the molecular emission linked to this source  is detected in the 
velocity interval [+2.7,+5.3] \kms. Both sources 1 and 3 are buried in the same molecular clump.} 
The detection of \hco\ towars source 3 also reveals the presence of dense molecular gas, which is linked to this source.

\subsection{Physical parameters of the molecular {  clump}}
\label{clumps_parameters}

From the $^{13}$CO(3$-$2) and C$^{18}$O(3$-$2) emission lines toward source 1 and 3, we determine the optical 
depth of each molecule ($\tau_{13}$ and $ \tau_ {18} $), excitation temperature ($T_{\rm ex}$), column 
density and mass of the molecular {  clump linked to both sources}. The excitation temperature was derived 
assuming LTE conditions and that the emission in the \tco\ line is optically {thick ($\tau_{\rm 13CO} >>$ 1). 
Using} the equations by \cite{buckle2010}, 
the excitation temperature result is 19.3 K.
{Assuming} that $T_{\rm ex}$ is the same for the two isotopologues, we estimate optical 
depths for C$^{18}$O and $^{13}$CO as 0.6 and 3.2, respectively.  

From the C$^{18}$O column density we estimate the mass as
\begin{equation}
 M_{\rm H2} = [H_{2}/C^{18}O] \ \mu_{\rm m} \ m_{\text{H}} \ A \ N(C^{18}O),
\end{equation}
\noindent
where [H$_{2}$]/[C$^{18}$O] =  6$\times$10$^{6}$ \citep{frerkin1982} is the 
{  molecular hydrogen-carbon monoxide}  abundance and  $A$ is the area of the clump. From the \dco\ contours 
of Fig. \ref{molec_cores_1}, we adopt an equivalent radius of 50\arcsec\ {  (or 0.45 pc at 2.0 kpc) for the 
molecular clump linked to the sources.} {Adopting} values for $\Delta \rm V_{\rm 13}$ and $\Delta \rm V_{\rm 18}$ 
{  of} 2.25 and 1.5 \kms, {respectively, we estimate} an H$_{2}$ column density of 1.4$\times$10$^{22}$ \cmdos\ 
and a mass of 310 \msun. A volume density $n_{H_{2}}  \sim$ 1.2$\times$10$^{4}$ \cmtres\ {  is derived by 
distributing the molecular mass within a sphere of 0.45 pc in radius.} Masses and H$_2$ volume densities are 
within the values derived for clumps in other regions of the Galaxy.

The detection of HCO$^+$ and HCN toward sources 1 and 3  indicates regions with high ambient densities with 
values of up to the critical density of the HCO$^+$ line ($\simeq$ 3$\times$10$^6$ \cmtres).

\subsection{Virial mass}

The virial  mass of the clump can be obtained from the \dco\ line. Considering only gravitational and internal 
pressure (i.e., neglecting support of magnetic fields, internal heating sources, or external pressure)  and  assuming a spherically 
symmetric cloud with an $r^{-2}$ density distribution, the virialized  molecular mass of the whole clump, 
$M_{\rm vir}$, can  be estimated from
$\quad M_{\rm vir}\,=\,126\ R_{\rm eff}\ (\Delta  {\rm v}_{\rm cl})^2$ 
\citep{MacLaren_1988}. In this expression, $R_{\rm eff}$ = $\sqrt{A_{\rm cl}/ \pi}$ is the effective radius in 
parsecs, $A_{\rm cl}$ is the area of the clump, and $\Delta {\rm v}_{\rm cl}$  is the width of the composite 
spectrum, which is defined in the same manner as for $\Delta{\rm v}^{18}$ in Fig.~\ref{molec_cores_1}. We find that the 
virial mass  is $\sim$ 135 M$_{\sun}$, which is much less than the LTE mass (310 M$_{\sun}$); this suggests that the clump
is collapsing. This result is compatible with the presence of embedded YSOs.

\section{Molecular outflows}
\label{Molecular_Outflows}

\begin{figure}[ht!]
\centering
\includegraphics[width=6.5cm]{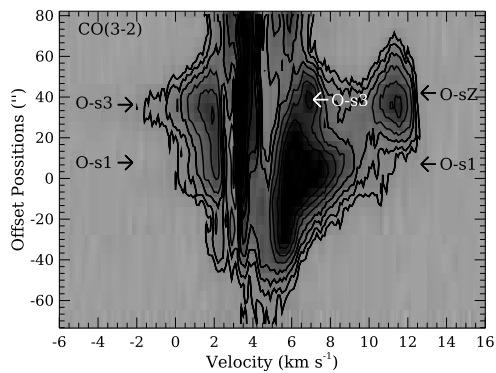}
\caption{Position-velocity diagrams of the $^{12}$CO($3-$2) molecular emission along a line from the 
northeast to southwest direction through the sources 1 and 3. Contour levels correspond 
to 1.5, 2.3, 3.5, 5.0, 7.0, 9.0, 11.0, and 14.0 K. Emissions labeled as O-s1 and O-s3 would be the redshifted and 
blueshifted lobes associated with sources 1 and 3, respectively.}
\label{pos_vel_CO_f1}
\end{figure}

Figure \ref{pos_vel_CO_f1} shows the position-velocity diagrams of the \co\ line emission along a cut from 
the northeast to southwest direction passing through the positions of sources 1 and 3, where the offset position 
0.0$''$ corresponds to the spectra through the center of source 1. The brightest 
emission, with velocities in the range $\sim$ $+$2.7 $< {\rm v} <$ $+$5.0 \kms, within the outer dashed lines 
in Fig.~\ref{molec_cores_1}, corresponds to the clump. Figure \ref{pos_vel_CO_f1} reveals molecular gas with 
velocities that are larger than those of the molecular turbulence ($\lesssim$ $+$2.5 \kms) outside the mentioned velocity 
range. The cut shows a prominent redshifted emission from $\sim$ $+$5.5 to $+$10 km\,s$^{-1}$ , 
labeled as O-s1, and a less notorious emission from $\sim$\,$+$6.0 to $+$8.5 \kms, indicated as O-s3 in white. 
This emission is overlapped by a cloud (O-sZ) showed in the 
velocity range $\sim$ $+$9.5\,$<\,{\rm v}\,<$\,$+$13.0 \kms. However, this particular emission might also be 
considered as a {  small cloud} that is extended from $\sim$\,$+$6.0 to $+$13.0\,\kms and connected 
to source 3. 
For blueshifted 
velocities, the extended emission is clearly detectable from $\sim$\,$-$1.5\,\,to\,\,$+$2.7\,\kms and from 0.0 to 
$+$2.7 \kms. These emissions would be associated with the redshifted emission O-s3 and O-s1, respectively.

The outflowing gas is generally shown by an optically thick line (such us $^{12}$CO), while the central clump 
is detected in an optically thin line (such as C$^{18}$O). Bearing in mind the \co\ and \dco\  lines (Fig. 2) ,
we take into account that, {for source 1,} the blue and red wings are defined in the velocity intervals [0.0,+2.7] and [+5.5,+10] \kms, 
respectively, while for source 3, these intervals are [--1.5,+2.7] and [+6.0,+13] \kms. We believe that 
molecular material with velocities outside the two dashed lines can be explained as molecular outflows 
originated in sources 1 and 3.

\begin{figure}[ht!]
\centering
\includegraphics[width=7.0cm]{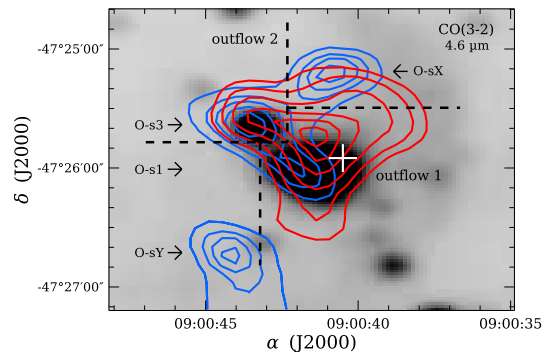}
\caption{Integrated map emission of the \co\ line wing profiles toward sources 1 and 3 superposed onto the 4.6 $\mu$m 
WISE image (grayscale). Light blue contours correspond to 7, 10, 12, 14, and 16 K\,\kms,\, while red contours correspond to 
18, 24, 32, 40, 45, and 48 K\,\kms.}
\label{outflows_lobes}
\end{figure}

In Figure 4 we show the integrated emission of the CO line within the redshifted and blueshifted
velocity intervals indicated above. Four structures can be distinguished in blueshifted velocities,
two of these are over sources 1 and 3 (O-s1 and O-s3) and the other two are labeled as O-sX and
O-sY. In redshifted velocities, we find extended molecular emission peaking on sources 1 and
toward source 3. On the contrary, no clear redshifted velocity component is associated with O-sX
and/or O-sY. 
{O-sX might be linked to the blueshifted outflows of one of the sources (see Fig. \ref{outflows_lobes}), since 
emission with the velocity range of this component is detected toward both sources in $^{12}$CO and $^{13}$CO.}

%

\begin{table}[!ht]
\centering
{\small
\caption{Parameters of the molecular outflows from the \co\ line}
\begin{tabular}{lccccc}
\hline
\hline
Source &  Lobe & $\tau_{12}$ & $\phi$   & $\Delta {\rm v}$ &   $M$          \\
       &       &  &        (arcsec) & (km\,s$^{-1}$)  &  ($M_{\sun}$) \\
\hline
     1 & O-s1 blue & -- & 30   & 4.5    & {  14}  \\
       & O-s1 red  & {  18} &24    &  2.4   & {  4.8} \\ 
\hline
     3 & O-s3 blue & -- & 33   &  2.5   & {  3.8}  \\
       & O-s3 red  & {  10} & 25   & 1.8    & {  1.7}  \\ 
\hline
\hline
\end{tabular}
\label{param_outflows}}
\end{table}

{  Adopting an abundance of [$^{12}$CO]/[$^{13}$CO] = 74 for a galactocentric distance of 8.8 kpc \citep{Wilson_Rood_1994}}, 
we estimate the opacities for the gas in the molecular outflows. 
Since the blueshifted lobes are not detected at \tco, 
we adopt the  same optical depths as for the redshifted lobes. 
Excitation temperatures are $\sim$ 15 K for both outflows. 
Table \ref{param_outflows} summarizes the optical depths for redshifted lobes, the effective 
diameters $\phi$, the velocity width of each lobe, and the masses of the blueshifted and redshifted lobes. 
{  The detection of outflows is compatible with previous findings by \cite{Wouterloot_1989}.}

\section{Summary}
\label{Conclusions}

IRAS 08589$-$4714 was observed in five molecular lines with the APEX telescope to characterize the molecular environment. 
An area of $\sim$\,150$''$ $\times$ 150$''$,  centered on the IRAS source position, was covered in the (3$-$2) transition of $^{12}$CO, 
$^{13}$CO, C$^{18}$O, HCO$^{+}$, and HCN lines. 
A search for candidate YSOs in the WISE database allowed us to identify three IR point sources with characteristics of 
Class I/II objects according to the criteria by \cite{Koenig_2012} within the surveyed region.

The molecular line profiles of  $^{12}$CO, $^{13}$CO, and C$^{18}$O, show  multiple velocity components and strong broadening 
effects toward sources 1 and 3. The spatial distribution of the CO emission shows the presence of a molecular clump of 0.45 pc 
in radius with mass and H$_2$ volume density of 310 $M_{\sun}$ and 1.4$\times 10^{4}$ cm$^{-3}$, respectively. The comparison between the LTE 
and virial mass indicates that the clump is collapsing. The HCO$^+$ and HCN spectra reveal  molecular overdensities that are coincident 
with sources 1 and 3. Finally, we detect two possible outflows associated with each source. 

\begin{acknowledgements}
  We thank Ramiro Franco for coordinating the observations with the APEX telescope. H. P. S. acknowledges financial support 
  from a fellowship from CONICET.
This project was partially financed by CONICET of Argentina under projects PIP 00356, and PIP 00107 and from UNLP, projects 
PPID092, PPID/G002, and 11/G120. M.R. wishes to acknowledge support from CONICYT (CHILE) through FONDECYT grant No1140839. 
\end{acknowledgements}

\bibliographystyle{aa} 
\bibliography{biblio} 

\end{document}